\documentclass[oneside]{article}
\usepackage{amssymb}
\usepackage{mathtools}
\usepackage{xcolor}
\usepackage{enumitem}
\usepackage{graphicx}
\usepackage{pdflscape}
\usepackage{url}
\usepackage{hyperref}
\usepackage{float}
\usepackage{natbib}
\usepackage{setspace}
\usepackage{fancyhdr}

\begin{document}
\pagestyle{fancy}
\thispagestyle{empty}
\fancyhead[r]{Preprint submitted to International Journal of Multiphase Flow}

{\Large\noindent\textbf{Uncertainties in Measurements of Bubbly Flows Using Phase-Detection Probes}}\newline

\noindent Matthias B\"urgler$^1$, Daniel Valero$^{2,3}$, Benjamin Hohermuth$^1$, Robert M. Boes$^1$, David F. Vetsch$^1$ \newline\newline
\noindent$^1$Laboratory of Hydraulics, Hydrology and Glaciology (VAW), ETH Zurich, Hönggerbergring 26, 8093 Zurich, Switzerland.\\%
\noindent$^2$Institute for Water and River Basin Development (IWG), Karlsruhe Institute of Technology (KIT), Engesserstraße 22, D-76131 Karlsruhe, Germany.\\%
\noindent$^3$Water Resources and Ecosystems department, IHE Delft, Westvest 7, 2611AX Delft, the Netherlands.\\

\noindent\textbf{Correspondance:} Matthias B\"urgler (buergler@vaw.baug.ethz.ch)\\

\noindent\textbf{Abstract.} The analysis of bubbly two-phase flows is challenging due to their turbulent nature and the need for intrusive phase-detection probes. However, accurately characterizing these flows is crucial for safely designing critical infrastructure such as dams and their appurtenant structures.
The combination of dual-tip intrusive phase-detection probes with advanced signal processing algorithms enables the assessment of pseudo-instantaneous 1-D velocity time series; for which
 the limitations are not fully fathomed.
In this investigation, we theoretically define four major sources of error, which we quantify using synthetically generated turbulent time series, coupled with the simulated response of a phase detection probe. 
Our findings show that typical high-velocity flows in hydraulic structures hold up to 15\% error in the mean velocity estimations and up to 35\% error in the turbulence intensity estimations for the most critical conditions, typically occurring in the proximity of the wall. 
Based on thousands of simulations, our study provides a novel data-driven tool for the estimation of these baseline errors (bias and uncertainties) in real-word phase-detection probe measurements.
\newline
\noindent\textbf{Keywords:} Aerated flow, Adaptive Window Cross-Correlation, Conductivity probe, Fibre optic probe, Signal processing, Two-phase flow turbulence

\section{Introduction}

Bubbly flows are relevant in many engineering applications, such as nuclear power reactors \citep{Vernier1968}, industrial pipe flows \citep{Beggs1973} or self-aerated flood spillways of dams \citep{Straub1958}. In the latter, flows can reach very high Reynolds numbers \citep{Cain1978, Hohermuth2021a}, with very high energy levels and severe consequences when poorly designed \citep{Vahedifard2017, Wahl2019}. Those flows require careful assessment of their properties and are typically studied under laboratory conditions, due to difficulty of access and violent flow conditions in prototypes.\\

Measuring self-aerated flow properties is complicated since classic instrumentation is typically limited to air concentrations ($c$) smaller than a few percents \citep{Chanson2013}. Pitot tubes \citep{Matos2002} or Acoustic Doppler Velocimetry (ADV) \citep{Nikora1998, Matos2002} are adversely affected by the presence of air bubbles, especially at higher concentrations. Other instrumentation such as Laser Doppler Velocimetry (LDV) \citep{Jones1976} and Particle Image Velocimetry (PIV) \citep{Hornung1995}, or photography are typically limited to just a few centimeters close to side walls due to limited visual accessibility \citep{Bung2016, Zhang2018, Kramer2020b}.\\

The most commonly used instrumentation -- applicable over a wide range of void fractions -- are dual-tip phase-detection probes, comprising conductivity (CP) or fibre optic (OF) probes. The former synchronously samples the flow resistivity (different for air and water), while the latter does the same for the light refraction, likewise different for air and water. Both probe types lead to very similar results, with OF probes offering improved detectability for smaller bubble sizes at the cost of more fragile probes \citep{Felder2017b, Felder2019}. Over the last few decades, dual-tip phase-detection probes with varying probe geometries have been applied in a large range of flow conditions by numerous studies, a selection of which is given in Table~\ref{tab:parameter_literature}.\\

With both probe types, the cross-correlation of sampled signals can yield the time-averaged velocity estimation through:
\begin{equation}
    \overline{u} = \Delta x / \tau
    \label{eq:velocity}
\end{equation}
with $\Delta x$ being the streamwise tip separation and $\tau$ the time-lag corresponding to the maximum cross-correlation between leading and trailing tip signal \citep[e.g.][]{Chanson2013}.\\

Recently, Eq.~\ref{eq:velocity} has been extended to allow multiple, pseudo-instantaneous velocity estimations by prior segmentation of the signal into small windows containing a pre-defined number of particles ($N_p$). This so-called Adaptive-Window Cross-Correlation (AWCC) algorithm \citep{Kramer2019b,Kramer2020a}, has enabled improved turbulence estimations compared to full signal cross-correlation (FSCC) based techniques \citep{Kramer2021}. This is supported by a recent comparison of the FSCC and AWCC approaches to a manual bubble identification (MBI) approach for estimating time-averaged velocities and turbulence intensities in a hydraulic jump \citep{Wang2021b}. In the lower jet-shear region, \citet{Wang2021b} reported median relative errors of 6\% for mean velocities and underestimations of turbulence intensities by 32\% when estimated with AWCC ($N_p$~=~15), compared to the MBI approach. The FSCC approach resulted comparable mean velocity errors of 4\%, but substantially larger turbulence intensity errors if 590\%. \\

Considering the substantial amount of manual data processing associated with MBI, AWCC presents an expedient alternative for mean velocity and turbulent intensity estimations in highly aerated flows with dual-tip phase-detection probes. However, careful assessment of biases, errors and limitations is required to shed light on measurement uncertainties and baseline errors.\\

\citet{Hohermuth2021b} and \citet{Pagliara2023} found that estimated velocities might be underestimated due to surface-tension effects that can cause a deceleration of bubbles during the interaction with the probe tips, with typical errors in the range of 5-10\%. This bias can be reduced to errors below 2\% with a proposed correction scheme \citep{Hohermuth2021b, Pagliara2023}. 
Additional uncertainties may arise from an implicit smoothing of the true turbulent velocity time series introduced by the AWCC algorithm by processing the signal in segments containing $N_p$ particles, as addressed by \citet{Kramer2020a} and later discussed by \citet{Chanson2020} and \citet{Kramer2021}. The magnitude of the smoothing depends on the time scale of the turbulent fluctuations $\mathcal{T}$, the bubble detection frequency, and the number of bubbles per window. While reducing $N_p$ reduces the smoothing effect, it also increases the probability of wrongly matching phase-change events \citep{Kramer2021}. To overcome the smoothing effect for turbulence estimations, \citet{Kramer2020a} suggest a method to extrapolate turbulence intensities to single particle statistics. However, this method requires calibration to the given flow conditions. \\

Additional uncertainties emerge from the one-dimensional flow assumption, despite frequently measuring in flows characterized by three-dimensional turbulence. Based on Monte Carlo simulations, \citet{Wang2018} found non-negligible uncertainty of dual-tip probes for bubbly flow measurement related to three-dimensional turbulence, probe geometry, and bubble properties (size, aspect ratio, and orientation). Since \citet{Wang2018} compared velocity estimations based on single-event detection to ground truth velocities, those results are not directly transferable to velocity estimations obtained with the AWCC algorithm, which is a well established methodology for high-velocity, highly aerated spillway flows.\\

In this study, we produce synthetic signals by using synthetic turbulent velocity time series generated based on the Langevin equation, which features power-law spectra with -5/3 slope, and adapted to laboratory-scale spillway flow properties sampled in \citet{Buergler2023a, Buergler2023b} with LDA, while covering a wide range of flow conditions encompassing also prototype measurements. We simulate the response of phase detection probes by reproducing synthetic voltage signals, which can thereafter be processed with the AWCC algorithm. By doing so, we have full access to the generated signal properties and the sampled \textit{true} flow properties, both in terms of air concentrations and turbulent velocity statistics. This allows to investigate the effect of flow properties and instrumentation characteristics on measurement uncertainty and to identify key mechanism resulting in bias or random errors of estimated flow properties. Building on dozens of thousands of synthetic turbulence and instrumentation response simulations, we expand on previous simpler error analysis \citep{Kramer2019b, Kramer2020a, Chanson2020, Kramer2021, Wang2018, Wang2021b} and provide a novel tool to estimate uncertainty of future air--water flow studies.

\begin{landscape}
\begin{table}[ht]
    \setlength{\abovecaptionskip}{0pt}%
    \setlength{\belowcaptionskip}{10pt}%
    \caption{Selected studies with application of dual-tip phase-detection probes in aerated, high-velocity chute or tunnel flows, including key parameters of flow conditions (time-averaged velocity $\overline{u}$, turbulence intensity $\mathrm{T}_u$) and of air-water flow instrumentation (streamwise and lateral tip separations $\Delta x$ and $\Delta y$, and sampling frequency $f_s$).}
    \centering
    \begin{tabular}{lllllll}
     \hline
     Reference &  $\overline{u}$ & $\mathrm{T}_u$ & $\Delta x$ & $\Delta y$ & $f_s$ & Comment\\
      & $\mathrm{[m/s]}$ & [-] & $\mathrm{[mm]}$ & $\mathrm{[mm]}$ & $\mathrm{[kHz]}$ & \\
     \hline
     \citet{Cain1978}   & 12-22 & n/a & 101.6 & 0 & 8 & Prot.; smooth spillway \\
     \citet{Chanson2002}   & 1-3.5 & n/a & 8 & 0  & 20 & Exp.; stepped spillway \\
     \citet{Boes2003}   &  up to 7 & n/a & 1.2-2.1 & 0  & $\leq$ 800 & Exp.; stepped spillway \\     \citet{Felder2015}   & n/a & 1.2 & 7.2 & 2.1  & 1-40 & Exp.; stepped spillway \\
     \citet{Bung2011}   & $\approx$~3.5 & n/a & 5.1 & 1 & 25 & Exp.; stepped spillway \\     
     \citet{Meireles2012}   & $\approx$~4.5 & n/a & n/a & n/a  & n/a & Exp.; stepped spillway \\     
     \citet{Felder2016}   &$\approx$ 2-4 & n/a & 7.2 & 2.1  & 20 & Exp.; stepped spillway \\
     \citet{Zhang2016}   & $\approx$~3.5 & n/a & n/a & n/a  & 20 & Exp.; stepped spillway \\     
     \citet{Felder2017a}   & ~2-5 & 1-1.4 & 7.2 & 0  & 20 & Exp.; stepped spillway \\
    \citet{Felder2017b}   & up to 7.5 & 0.8-4 & 1.77-5.17 & 0-1.05  & 500 & Exp.; stepped spillway \\
     \citet{Felder2019}   & 4-22 & n/a & 5.06 & 0.97  & 500 & Exp.; tunnel chute \\
     \citet{Kramer2019b}   & 2-6 & 0.1-0.5 & 4.7 & n/a  & 20 & Exp.; stepped spillway \\
     \citet{Kramer2020b}   & 2-12 & 0.05-0.25 & 5.06 & 0.97  & n/a & Exp.; tunnel chute \\
     \citet{Hohermuth2021a}   & 23-38 & n/a & 2 & n/a & 500 & Prot.; tunnel chute \\
     \citet{Hohermuth2021b}   & 0.2-12 & n/a & 4.07-5.36 & 1-1.41  & 500 & Exp.; tunnel chute \\
     \citet{Nina2022}   & ~4 & 0.25-2 & 6.2 & 1.35 & 500 & Exp.; stepped spillway \\
     \citet{Cui2022}   & up to $\approx$ 4 & n/a & 8.104 & n/a  & 40 & Exp.; grass-line spillway\\
     \citet{Felder2023}   & ~4 & n/a & 4.85 & 1.1  & 20 & Exp.; micro-rough spillway\\
     
     \hline
    \end{tabular}
    \label{tab:parameter_literature}
\end{table}
\end{landscape}

\fancyhead[r]{Preprint submitted to International Journal of Multiphase Flow}

\section{Methods}\label{sec:methods}

Stochastic velocity time series are well suited to investigate accuracy and limitations of air-water flow instrumentation, such as phase-detection probes \citep[e.g.][]{Bung2017, Valero2019, Kramer2019a, Kramer2019b, Buergler2022}. In this study, we develop an open-source software for simulating phase-detection probe measurements within bubbly flows \citep{Buergler2024b}. The software allows to generate stochastic velocity time series representative of turbulent flows.
Randomly generated bubbles are superposed to this velocity signal, and are thereafter tracked while registering any intersection with the tips of a virtual dual-tip phase detection probe, which is recorded in the form of synthetic voltage signals.\\

\begin{figure}[ht]
\centering\includegraphics{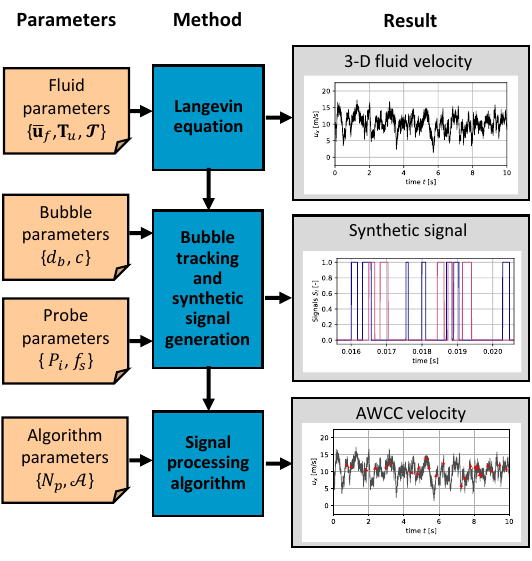} \caption{Workflow to generate stochastic velocity time series and synthetic phase-detection probe signals, as well as signal post-processing. Input parameters are the mean (time-averaged) fluid velocity $\mathbf{\overline{u}}_f$, turbulence intensity $\mathbf{\mathrm{T}}_u$, integral time scale $\mathcal{T}$, bubble diameter $d_b$, air concentration $c$, probe tip locations $\mathbf{P}_i$, sampling frequency $f_s$, number of particles per window $N_p$, and cross-correlation based filtering threshold $\mathcal{A}$.}\label{fig:workflow}
\end{figure}

Bubble-probe interaction effects are not included in this methodology (see \citet{Hohermuth2021b}, and \citet{Pagliara2023}), at the expenses of allowing results independent of the dual-tip probe diameter; or, in other words, assuming that the tip diameter is small enough compared to the bubble size. The synthetic signals are then post-processed using the AWCC algorithm of \cite{Kramer2019b}, including latest filtering methods by \cite{Kramer2020a}, to recover various flow properties, such as the pseudo-instantaneous velocity time series, the time-averaged velocity, the turbulence intensity, or the air concentration. This workflow is depicted in Figure~\ref{fig:workflow}. For the workflow presented herein, we assume that the main flow direction is aligned with the $x$ coordinate axis, the $y$ coordinate axis represents the transversal and $z$ the wall-normal direction. \\

\subsection{Stochastic Velocity Time Series}

The stochastic 3-D velocity time series of the fluid $u_{i,f}(t)$ for $i$=$x,y,z$ is generated with the Langevin equation in finite difference form \citep{Pope2000}:
\begin{equation}
    u_{i,f}(t+\Delta t_r) = u_{i,f}(t) 
    - \left(u_{i,f}(t) -\overline{u}_{i,real}\right)  \frac{\Delta t_r}{\mathcal{T}_{i,real}} + u_{rms,i,real}\sqrt{\frac{2\Delta t_r}{\mathcal{T}_{i,real}}}\xi_i(t),
    \label{eg:langevin}
\end{equation}
where $u_{rms,i,real}$ = $\overline{u}_{i,real}\mathrm{T}_{u,i,real}$ is the root mean square (RMS) velocity fluctuation, $\overline{u}_{i,real}$ the mean (time-averaged) velocity, $\mathrm{T}_{u,i,real}$ the turbulence intensity, $\mathcal{T}_{i,real}$ the integral time scale,  and $\Delta t_r$ is the discrete time step. $\xi_i(t)\sim\mathcal{N}$(0,1) are the respective components of a standardized multi-variate Gaussian random variable with zero mean. The diagonal elements of the covariance matrix  are equal to 1 and the off-diagonal elements are zero with exception of $\rho_{xz,real}$~=~--0.45, taking into account the velocity covariance typically defining the shear stress in boundary layer flows \citep{Pope2000}. 

\fancyhead[]{}
\fancyhead[r]{Preprint submitted to International Journal of Multiphase Flow}

\subsection{Synthetic Signal Generation}

We assume a control volume with edge lengths $L_i$~=~$\mathrm{max}(\Delta P_{i,j}) + 2d_b$, with $\mathrm{max}(\Delta P_{i,j})$ being the maximum distance in $i$-direction between the $j$-th probe tips, and $d_b$ being the bubble diameter. The virtual phase-detection probe is located at the center of the control volume. Then, a finite number of bubbles is defined as $N_b=f_b\cdot T$, where $T$ is the duration of the stochastic velocity time series and $f_b=1.5c_{real}\left|\overline{\mathbf{u}}_{real}\right|/d_{b,real}$ is the bubble count rate, with $\left|\overline{\mathbf{u}}_{real}\right|$ being the magnitude of the time averaged velocity.  
For each bubble $k$, an arrival time $t_{a,k}$ is defined assuming equal inter-arrival times $t_{iat}=T/N_b$.
\begin{equation}
    t_{a,k} = k\cdot t_{iat}
\end{equation}
At the arrival time $t_{a,k}$, bubble $k$ is then placed at random arrival location in the $y$-$z$ plane at the beginning of the control volume:
\begin{equation}
    \mathbf{x}_{k}(t_{a,k}) = \begin{pmatrix} -L_x/2 \\ \mathcal{U}_{[-L_y/2,L_y/2]} \\ \mathcal{U}_{[-L_z/2,L_z/2]}  \end{pmatrix}
\end{equation}

The trajectory of the $k$-th bubble inside the control volume is obtained by applying the first-order Euler method to the differential equation $u_{i,k}(t)$~=~$\partial x_{i,k}/\partial t$, yielding:
\begin{equation}
    \mathbf{x}_{k}(t+\Delta t_r) = \mathbf{x}_{k}(t) + \mathbf{u}_{k}(t)\Delta t_r.
    \label{eq:trajectory}
\end{equation}

Therefore, we assume that the slip ratio between the bubble velocity and fluid velocity $K = \mathbf{u}_{b,k}/\mathbf{u}_{f}$ is equal to one. This is a typical assumption for the prediction of velocities in aerated flows using phase-detection probes \citep[e.g.][]{Cain1978, Chanson1988, Hohermuth2021b}.\\

To take into account the finite sampling frequency $f_s$ of the virtual probe, the bubble trajectory time series $\mathbf{x}_{k}(t)$ is resampled to a time step $\Delta t_s=1/f_s$ by linear interpolation. For each time $t_s$ of the $k$-th bubble trajectory, we evaluate whether bubble $k$ is pierced by probe tip $j$ based on the inequality:
\begin{equation}
    \frac{(P_{x,j} - x_{x,k}(t_s))^2+(P_{y,j} - x_{y,k}(t_s))^2+(P_{z,j} - x_{z,k}(t_s))^2}{(d_b/2)^2} < 1.
    \label{eq:piercing_inequality}
\end{equation}

The synthetic signal takes values of 1, if the sensor $j$ lies inside the bubble (Eq.~\ref{eq:piercing_inequality} satisfied), and 0 otherwise.

\subsection{Signal Processing Algorithm}
The AWCC algorithm breaks the synthetic signal into $N_{\mathcal{W}}$ small segments, each segment covering a group of $N_p$ particles detected by the leading tip. Extending Eq.~\ref{eq:velocity}, the pseudo-instantaneous 1-D velocity of the $i$-th window is then estimated as:
\begin{equation}
    \left[u_{x,i,awcc}\right]_{t_{i}}^{t_{i}+T_{\mathcal{W},i}} = \Delta x / \tau_i,
\end{equation} \label{eq:u_x_awcc}
where $t_{i}$ is the start time of the $i$th window, $T_{\mathcal{W},i}$ is the window duration, $\Delta x$ is the streamwise tip separation, and $\tau_i$ is the travel time of particles within the $i$th window, estimated as $\tau_i = arg max(R_{12,i})$, where $R_{12,i}$ is the cross-correlation function between the leading and trailing tip signals contained in the $i$th window.

A window is rejected if \citep{Kramer2019b,Kramer2020a}:
\begin{equation}
    R_{12,i,max}/(SPR_i^2+1) < \mathcal{A}
    \label{eq:correlation_filter}
\end{equation}
where $R_{12,i,max}$ is the maximum cross-correlation coefficient, $SPR$ the secondary peak ratio of window $i$ and the parameter $\mathcal{A}~=~0.4$, as recommended by \citet{Kramer2020a}.
Further, the measured velocity time series are filtered iteratively until no more outliers are rejected, using a robust outlier cutoff (ROC) filter \citep{Valero2020} based on the median (MED) and the median absolute deviation from the median (MAD) as estimators of location and scale, based on the recommendations of \citet{Wahl2003}. Rejected values are replaced by NaN values (i.e., Not a Number), which do not contribute to statistical inference of turbulence properties.
To eliminate a sampling velocity bias due to more particles passing the probe at large velocities, window duration weighted mean velocity and root mean square velocity fluctuations are calculated as \citep{Buchhave1979}:
\begin{equation}
    \overline{u}_{x,awcc} = \frac{\sum_{i=1}^{N_{\mathcal{W}}}w_iu_{x,i,awcc}}{\sum_{i=1}^{N_{\mathcal{W}}}w_i},
    \label{eq:u_mean_awcc}
\end{equation}
and: 
\begin{equation}
    u_{rms,x,awcc} = \sqrt{\frac{\sum_{i=1}^{N_{\mathcal{W}}}w_i\left(u_{x,i,awcc}-\overline{u}_{x,awcc}\right)^2}{\sum_{i=1}^{N_{\mathcal{W}}}w_i}},
    \label{eq:u_rms_awcc}
\end{equation}
with weights $w_i = T_{\mathcal{W},i}$. Finally, the turbulence intensity is calculated as:
\begin{equation}
    \mathrm{T}_{u,x,awcc} = u_{rms,x,awcc}/\overline{u}_{x,awcc}.\\
    \label{eq:T_u_awcc}
\end{equation}

For mean velocity estimates, we calculate the 95\% confidence intervals as $\overline{u} \pm 1.96 \sqrt{u_{rms}^2/N}$, where $N$ is the number of pseudo-instantaneous velocity estimated obtained from the AWCC algorithm. Assuming the instantaneous velocity follows a normal distribution, we also obtain the 95\% confidence interval for root mean square velocity fluctuations as $u_{rms} \pm 1.96 \sqrt{u_{rms}^2/(2N)}$ \citep{Benedict1996}. Further, the 95\% confidence interval of turbulence intensity estimates were calculated as $\mathrm{T}_{u} \pm 1.96 \sqrt{\sigma_{\mathrm{T}}^2}$, with variance of the turbulence intensity $\sigma_{\mathrm{T}}^2$ obtained through error propagation as: 
\begin{equation}
    \sigma_{\mathrm{T}}^2 = \frac{u_{rms,x,awcc}}{\overline{u}_{x,awcc}}\left[ \left(\frac{\sqrt{u_{rms}^2/N}}{\overline{u}} \right)^2 + \left(\frac{\sqrt{u_{rms}^2/(2N)}}{u_{rms}} \right)^2 \right].\\
    \label{eq:CI_T_u_awcc}
\end{equation}

One of the effects we investigate in this research is the window-averaging effect introduced by estimating the pseudo-instantaneous velocity for a group of $N_p$ particles, which pass the probe during a time window $T_{\mathcal{W},i}$. For this, we calculate the pseudo-instantaneous velocities for each window $i$, by time-averaging the stochastic velocity time series over duration of the $i$th window:
\begin{equation}
    \left[u_{x,i,f}\right]_{t_{i}}^{t_{i}+T_{\mathcal{W},i}} = \int_{t=t_{i}}^{t_{i}+T_{\mathcal{W},i}} u_{f}(t) \, \mathrm{d}t
    \label{eq:u_fluid_window}
\end{equation}

For conciseness, we denote $\left[u_{x,i,f}\right]_{t_{i}}^{t_{i}+T_{\mathcal{W},i}}$ as $\langle u_{x,f}\rangle_{\mathcal{W},i}$ hereinafter.

\subsection{Stochastic signals verification: velocity and phase functions}

The workflow was verified by comparing velocity time series, measured with a Laser Doppler Anemometer (LDA), to stochastic velocity time series with equivalent mean and root-mean square velocity fluctuations. The velocity time series were sampled in the non-aerated region of a laboratory scale spillway chute, at relative invert normal distances $z/h$ of 0.03, 0.21 and 0.41, respectively, where $h$ is the flow depth. The spillway is characterized by an inclination angle of 50$^\circ$, and an invert roughness of $k_s$~=~0.74~mm. Further details on the experimental setup can be found in \citet{Buergler2023b}.

The velocity time series were recorded for 120~s at a specific flow rate of 0.2~m$^2$s$^{-1}$ at 3.67~m downstream of the broad-crested weir with mean sampling rates of 924~Hz, 1310~Hz, and 1525~Hz, respectively. The mean velocities, (turbulent intensities), and [integral time scales] are 4.22~ms$^{-1}$, 5.71~m$^2$s$^{-1}$, and 6.25~m$^2$s$^{-1}$, (0.20, 0.12, and 0.09), and [0.036~s, 0.13~s, and 0.24~s], respectively.\\

\begin{figure}[!ht]
\makebox[\textwidth][c]{\includegraphics{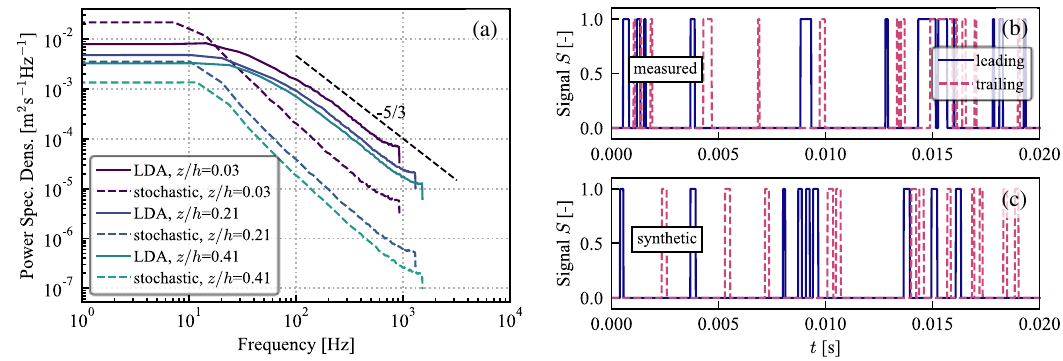}}
\caption{Power spectral density of three Laser Doppler Anemometry (LDA) velocity time series and stochastic velocity time series with the same mean velocity, turbulent intensity and integral time scale (a). Measured and binarized signals in a self-aerated boundary layer (b) and synthetically generated signals for comparable flow conditions (c).}
\label{fig:verification}
\end{figure}

For the sake of comparison, three stochastic velocity time series were generated with equivalent flow properties and duration.
The power spectral densities obtained with the Welch's method \citep{Welch1967} for the measured and stochastic velocity time series are compared in Figure~\ref{fig:verification}\textit{a}. Therefore, the measured LDA time series were adjusted for irregular sampling by zero order interpolation (sample-and-hold) \citep[e.g.][]{Velte2014}. All spectra exhibit a -5/3 decay. For the measured time series, the decay begins at slightly larger frequencies and approaches the -5/3 slope slower than the stochastic time series. It should be noted that additional noise in the LDA signals (filtered) may lead to certain deviations when compared to the stochastic signals. 
Nevertheless, the stochastic time series reproduce the essential characteristics of turbulent flow, while satisfying the same mean and turbulence intensity, as well as the velocity covariance in the $x-z$ plane.\\

For further verification of the proposed methodology, we compare synthetic signals to measured, binarized phase-detection probe signals recorded in the aerated region of the previously described experimental setting. The signals were recorded at a location $x$~=~8.17~m downstream of the spillway crest. Synthetic signals were generated for the same flow conditions and instrumentation characteristics. For more detail we refer again to \citet{Buergler2023b}.
The visual comparison of the measured and synthetic signals in Figure~\ref{fig:verification}\textit{b} and \textit{c}, respectively, reveals fundamental similitude, whereas air concentrations are both equivalent in the physical data and the generated stochastic series.

\section{Key mechanisms defining the baseline errors}
\label{sec:theory}

In this section, we theoretically address the key mechanisms defining the baseline errors occurring during the estimation of flow properties in bubbly turbulent flows.
Before the systematic quantification of these errors, we elaborate on four main mechanisms, that may result in a bias and/or random errors when sampling flow properties in aerated, bubbly flows with phase-detection probes, namely:

\begin{enumerate}[label={\it E\arabic*}]
    \item Selective sampling bias 
    \item Velocity bias 
    \item Decorrelation bias 
    \item Averaging effect due to pseudo-instantaneous velocity estimations
\end{enumerate}

While \textit{E1} to \textit{E3} are primarily caused by 3-D turbulence, \textit{E4} is a characteristic of the AWCC processing algorithm.

\subsubsection*{{\it E1:} Selective sampling bias}
Conductivity probes are usually installed aligned with the mean flow. However, as a result of turbulence (3-D), individual bubbles move along trajectories at an angle relative to the probe orientation. For the sake of simplicity, in the following we only consider turbulent fluctuations in the $x$-$y$-plane. Further, we assume spherically-shaped bubbles with diameter $d_b$ and a constant bubble velocity $\mathbf{u}_b$~=~$(u_{x,b},u_{y,b})$ during the interaction with the probe ($\Delta x / u_{x,b} \ll \mathcal{T}$). For this simplified case, the angle of bubble movement may be defined as $\theta$~=~$\tan^{-1}(u_{y,b}/u_{x,b})$, where $u_{x,b}$ and $u_{y,b}$ are the instantaneous bubble velocities in $x$- and $y$-directions, respectively.\\

\begin{figure}[!ht]
\makebox[\textwidth][c]{\includegraphics{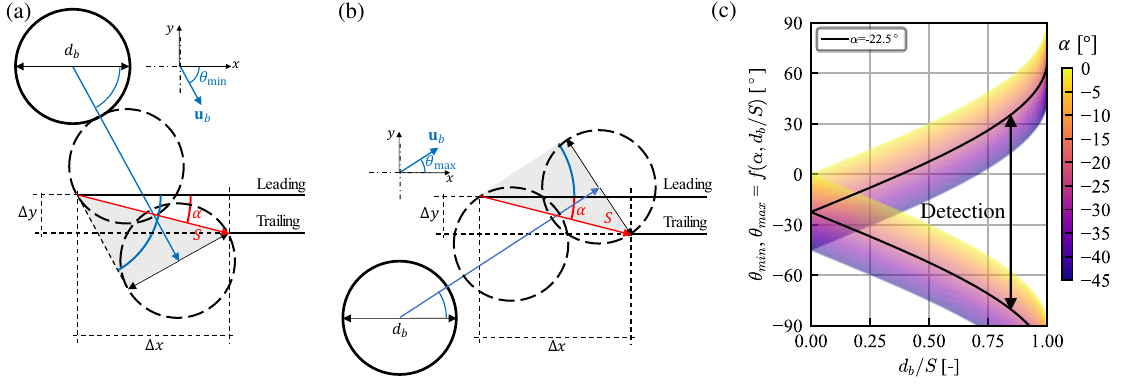}}
\centering 
\caption{Identification of the: (a) minimum ($\theta_{min}$) and (b) maximum ($\theta_{max}$) angles of detection for a bubble with velocity $\mathbf{u}_b$, as a function of $\alpha$ (tip-to-tip axis inclination), tip separation ($S$) and bubble diameter ($d_b$). (c) Mapping of the detection regions.}
\label{fig:detection_cones}
\end{figure}

For a given streamwise and transverse tip separation ($\Delta x$ and $\Delta y$, respectively), there is a limiting bubble trajectory angle for which the bubble can be detected by both probe tips. These minimum and maximum bubble trajectory angles ($\theta_{min}$ and $\theta_{max}$, respectively), are illustrated in Figure~\ref{fig:detection_cones}a-b. For the shaded triangles in Figure~\ref{fig:detection_cones}a and b it is possible to write:
\begin{equation}
    \mathrm{sin}\left(\theta_{min}-\alpha\right) = \frac{d_b}{S},
    \label{eq:triangle_theta_min}
\end{equation}
and:
\begin{equation}
    \mathrm{sin}\left(\theta_{max}+\alpha\right) = \frac{d_b}{S},
    \label{eq:triangle_theta_max}
\end{equation}
where $\alpha$~=~$\tan^{-1}(\Delta y / \Delta x)$ is the angle between the leading and trailing probe tips and $S$~=~$\sqrt{\Delta x^2 + \Delta y^2}$ the distance between the two tips. Solving Eq.~\ref{eq:triangle_theta_min} and~\ref{eq:triangle_theta_max} for the minimum and maximum angles of detection $\theta_{min}$ and $\theta_{max}$ results in:
\begin{equation}
    \theta_{min} = \sin^{-1}\left(\frac{d_b}{S} \right) + \alpha
    \label{eq:theta_min}
\end{equation}
\begin{equation}
    \theta_{max} = \sin^{-1}\left(\frac{d_b}{S} \right) - \alpha.
    \label{eq:theta_max}
\end{equation}

Figure~\ref{fig:detection_cones}c illustrates the minimum and maximum detection angles as a function of $\alpha$ and the ratio between the bubble diameter and the tip separation distance $d_b/S$. It is important to highlight that the angles of detection are symmetric around $\alpha$ and not around 0.
This indicates that probes with a side-by-side design ($\left|\alpha\right|>0$) exhibit a selective sampling bias, potentially resulting in a selective sampling of bubbles and thereby in biased mean velocity or turbulence intensity estimations.

\subsubsection*{{\it E2:} Velocity bias due to transverse impacts}
\citet{Thorwarth2008} described how a non-parallel alignment of the probe tips to the main flow direction leads to a systematic overestimation of the mean velocity. Assuming a large bubble interface, idealised as a plane orthogonal to the bubble trajectory, the ratio $\chi$ between the estimated velocity $u_{x,awcc}$ and the true streamwise velocity $u_{x,b}$ is given as \citep{Thorwarth2008}:
\begin{equation}
    \chi = \frac{u_{x,awcc}}{u_{x,b}} = \frac{\Delta x/\tau}{L_x/\tau} = \frac{\cos\left(\alpha\right)}{\cos\left(\theta\right)\cos\left(\theta - \alpha\right)}.
    \label{eq:chi}
\end{equation}

\begin{figure}[!ht]
\makebox[\textwidth][c]{\includegraphics{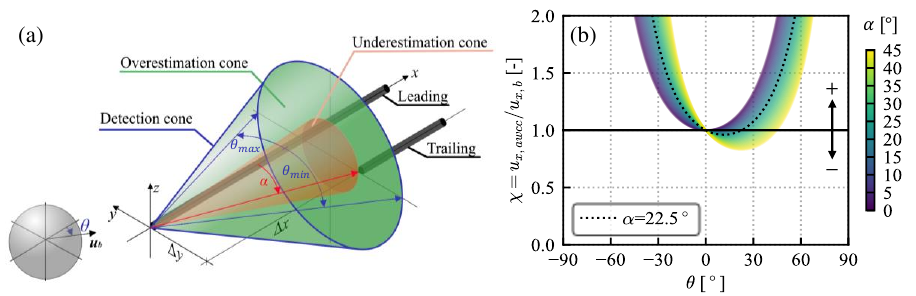}}%
\centering
\caption{(a) Bubble trajectory angles resulting under- or overestimation form cones in 3-D space. The minimum ($\theta_{min}$) and  maximum ($\theta_{max}$) angles of detection form a cone aligned with the tip-to-tip axis. (b) Ratio $\chi$ between the estimated velocity $u_{x,awcc}$ and the true streamwise velocity $u_{x,b}$ as a function of $\theta$ and $\alpha$.}
\label{fig:estimation_cones}
\end{figure}

Figure~\ref{fig:estimation_cones}b presents $\chi$ as a function of $\theta$ for a range of $-45 ^\circ \leq \alpha \leq 0^\circ$.
Although \citet{Thorwarth2008} focused on overestimations of mean velocities, Eq.~\ref{eq:chi} also showcases a mild underestimation ($\chi<1$) for $0^\circ > \theta > \alpha$. More interestingly, the underestimation appears to be bounded, while the overestimation is unbounded and strongly increases for $\theta \gg 0^\circ$ or $\theta \ll \alpha$. In 3-D space, the angles of under- or overestimation take the form of an under- and overestimation cones as depicted in Figure~\ref{fig:estimation_cones}\textit{a}. \\

This over- and underestimation, depending on the approach angle $\theta$, means that even for a constant streamwise velocity, but varying $\theta$ as a result of transverse turbulence fluctuations, a certain variance would be estimated, thus directly adding to an overestimation of turbulence.
Moreover, the over- and underestimation cones are not symmetrical around $\theta = 0^\circ$ for phase-detection probes with a side-by-side design ($\left|\alpha\right|>0^\circ$). This means that also mean velocity estimations are biased due to transverse turbulent fluctuations. \\

\subsubsection*{{\it E3:} Decorrelation bias}
\citet{Thorwarth2008} also pinpointed that increasing flow turbulence results in a larger number of bubbles that interact with only one of the two probe tips, leading to smaller maximum cross-correlation coefficients ($R_{12,max}$) during full signal cross-correlation based velocity estimations. For AWCC-based processing, this can further affect velocity estimations since windows with low $R_{12,i,max}$ are removed by the cross-correlation based filtering (Eq. \ref{eq:correlation_filter}). Based on the detection limits defined by Eq.~\ref{eq:theta_min} and \ref{eq:theta_max}, it is evident that this effect is exacerbated with decreasing ratio $d_b/S$, i.e., for smaller bubbles and larger tip-to-tip distance $S$.\\

Further, for large transverse and small streamwise velocity fluctuations, bubbles are more likely to hit only one of two tips as the angle $\theta = \tan^{-1}\left(u_{b,y}/u_{b,x}\right)$ with respect to the mean flow is larger than for large transverse and large streamwise velocity fluctuations. This results in smaller $R_{12,i,max}$ for windows during smaller streamwise velocity fluctuations and consequently a higher probability that those windows are removed by the correlation-based filtering. Ultimately, this results in an over-representation of windows with large streamwise velocities and directly overestimates the mean velocity, while also underestimating the turbulence intensity.

\subsubsection*{{\it E4:} Window averaging effect}
The AWCC algorithm breaks the recorded signals into windows, such that each window contains a number of $N_p$ bubbles.
The time duration of the window therefore depends on the interarrival time between bubbles or, on average terms, the bubble frequency $f_b$.

Assuming spherical bubbles, the latter can be estimated as $f_b = 1.5 \overline{u}c/d_b$ with $\overline{u}$ being the mean flow velocity, $c$ the local air concentration and $d_b$ the bubble diameter \citep{Crowe2005}. The average window duration is then defined as
\begin{equation}
    T_{\mathcal{W}} = \frac{N_p d_b}{1.5 \overline{u} c}.
    \label{eq:L_x}
\end{equation}
For large $T_{\mathcal{W}}$, there is an increasing risk of undersampling the real instantaneous velocity time series. If the flow has a small integral time scale $\mathcal{T}_x \ll T_{\mathcal{W}}$, a segment will cover several fluctuations, therefore smoothing the natural fluctuations of the flow. In simpler terms, if the velocity fluctuates around the mean value with a time scale $\mathcal{T}_x$ of 1~s and the segment $T_{\mathcal{W}}$ covers 10~s, we will average over ten fluctuations/cycles within a pseudo-instantaneous velocity estimation. Conversely, if for the same sampling characteristics, our flow time scale is 100 s, then the AWCC would be able to obtain 10 pseudo-instantaneous velocity estimations per turbulent cycle. Based on 1-D stochastic velocity time series and synthetic signals, \citet{Kramer2019b} found that turbulence intensity was underestimated by $\sim20 \%$ for $\mathcal{T}_x/T_{\mathcal{W}} \approx 1$, while the mean velocity estimations were not significantly affected.\\

\section{Results: Error quantification}
\label{sec:results}

\subsection{General remarks}
\label{sec:description_base_case}

In this section, we study how the key factors identified in Section~\ref{sec:theory} are manifested in self-aerated boundary layer flows and quantify their importance. For that purpose, we identify a wide range of flow properties, probe characteristics and post-processing parameters in literature (Table~\ref{tab:parameter_literature}). Relevant flow parameters that are not included in Table~\ref{tab:parameter_literature} are the velocity covariance $\rho_{xz}$, the integral time scale $\mathcal{T}$, and the bubble diameter $d_b$. The velocity covariance typically defining the shear stress
in boundary layer was assumed with a constant value -0.45 \citep{Pope2000}. For subcritical open-channel flow, \citet{Valero2022} observed integral time scales between $\mathcal{O}(-1)$~s to $\mathcal{O}(0)$~s. To account for the lack of experimental data on integral time scales in supercritical boundary layer flows, we investigate a wide range of integral time scales between 0.001~s and 1.0~s. Finally, we consider bubble diameters between 0.5~mm and 20~mm based on the range bubble size observed by \citep{Buergler2023a} in laboratory-scale spillway flows.\\

By default, we focus on a base case (identified in Table~\ref{tab:scenarios}), which represents a reasonable, representative flow situation and sampling procedure. Then, we systematically vary various parameters around the base case to aid the identification of key factors and to explain errors in estimated mean velocities and turbulent intensities. When necessary, we add further variables to identify the key mechanisms driving the errors in estimations.\\

\begin{landscape}
\begin{table}[!ht]
    \setlength{\abovecaptionskip}{0pt}%
    \setlength{\belowcaptionskip}{10pt}%
    \caption{Common range of flow conditions on spillway or tunnel chutes and parameters of phase-detection probes and selection of the base case for the quantification of baseline errors.}
    \centering
    \begin{tabular}{llllll}
     \hline
      Parameter&Symbol&Unit&Range&Base&Comment \\
     \hline
     \multicolumn{6}{l}{Flow conditions} \\
     \hline
     Mean velocity & $\overline{u}$ & [m/s] & 1-50 & 10 & Based on the range of flow velocities identified in Tab.~\ref{tab:parameter_literature}.\\
     Turbulent intensity & $\mathrm{T}$ & [m/s] & 0.0-0.35 & 0.25 &  Based on the range of turbulence intensities identified \\
      &  & & & & in Tab.~\ref{tab:parameter_literature}, but limited to $\mathrm{T} \leq 35\%$, since larger $\mathrm{T}_{u}$ result \\
      &  & & & & in negative instantaneous velocities, which is not physically \\
      &  & & & & meaningful for self-aerated flows on spillways.\\
     Velocity covariance & $\rho_{xz}$ & [-] & -0.45 & -0.45 & Velocity covariance typically defining the shear stress\\
      &  & & & & in boundary layer flows \citet{Pope2000}.\\
     Integral time scale & $\mathcal{T}$ & [s] & 0.001-1.0 & 0.1 & Based on integral time scales obtained from ADV \\
      &  & & & & measurements by \citet{Valero2022} and extended to \\
      &  & & & & account for smaller time scales.\\
     Void fraction & $c$ & [-] & 0.005-0.4 & 0.25 &  Based on the range of void fractions for which bubbly\\
      &  & & & &  flow can be expected. \\
     Bubble diameter & $d_{b}$ & [mm] & 0.5-20 & 3 & Based on the range bubble size identified observed   \\
      &  & & & & by \citet{Buergler2023a}. \\
     \hline
     \multicolumn{6}{l}{Phase-detection probe geometry and post-processing} \\
     \hline
     Streamwise separation & $\Delta x$ & [mm] & 0.5-10 & 5 & Based on the range identified in Tab.~\ref{tab:parameter_literature}.\\
     Transverse separation & $\Delta y$ & [mm] & 0-2 & 1 & Based on the range identified in Tab.~\ref{tab:parameter_literature}.\\
     Sampling frequency & $f_s$ & [kHz] & 10-2000 & 500 & Based on the range identified in Tab.~\ref{tab:parameter_literature}.\\
     \hline
    \end{tabular}
    \label{tab:scenarios}
\end{table}
\end{landscape}

For the following simulations, we sample for a duration of 300~seconds, thereby exceeding typical recommendations on sampling duration \citep[e.g.][]{Felder2015}.
Beside investigating the effect of flow factors and instrumentation characteristics on mean velocity and turbulent intensity estimations, we also investigated the effect on the estimation of void fractions. However, under our modelling assumptions, we found that the void fraction could be recovered from the measured signals with a negligible error ($\pm1$\%).\\

\subsection{Flow properties affecting the mean velocity estimation}
\label{sec:mean_vel_errors}

Estimations of mean (time-averaged) flow velocities based on phase-detection probe measurements are common practice and generally regarded as accurate. Using a cross-correlation based method, \citet{Thorwarth2008} found that mean velocities were overestimated by $\approx$~10\% in comparison to LDA-measurements, when the probe was aligned with the main flow direction. For highly turbulent conditions ($\mathrm{T_{u,x}}$~=~30\%), \citet{Kramer2019b} reported an overestimation of mean velocities up to 10\%, which was attributed to the fact that more particles pass the probe tips at higher velocities. This error could be reduced to negligible errors of $\sim$1\% by applying the window weighting scheme to calculate first and second order moments of the pseudo-instantaneous velocity time series \citep{Kramer2020a}, as used herein (Eqs.~\ref{eq:u_mean_awcc} and \ref{eq:u_rms_awcc}).\\

In the following, we test mean velocities between 2.5 to 40 m/s, around the base case of Table~\ref{tab:scenarios} and turbulence intensities between 0\% and 35\%. We investigate the effect of streamwise and transverse turbulence separately, by setting either $\mathrm{T}_{u,x}$ or $\mathrm{T}_{u,y}$ (and $\mathrm{T}_{u,z}$) to zero and systematically vary the other parameter. This analysis allowed us to identify turbulence as one of the main factors affecting the accuracy in the estimation of mean velocities.\\

\begin{figure}[ht]
\makebox[\textwidth][c]{\includegraphics{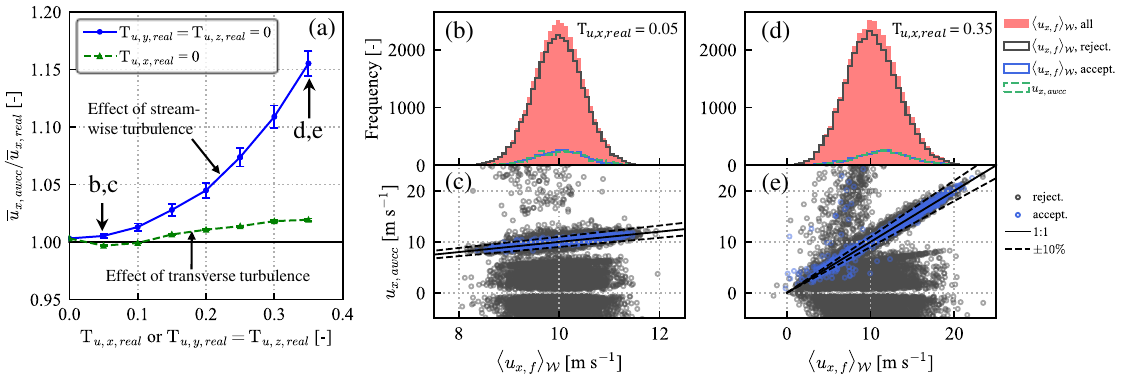}}
\caption{(a) Effect of streamwise ($\mathrm{T}_{u,x}$) and transverse ($\mathrm{T}_{u,y}$,$\mathrm{T}_{u,z}$) turbulence  on the mean velocity estimations. Frequency distributions of the window-averaged fluid velocity $\langle u_{x,f}\rangle_{\mathcal{W}}$ of `all', `rejected' or `accepted' windows and estimated pseudo-instantaneous velocity $u_{x,awcc}$ for $\overline{u}_{x,real}$~=~10~ms$^{-1}$ and $\mathrm{T}_{u,x,real}$ of (b) 5\% and (d) 35\%, respectively. Comparison of $u_{x,awcc}$ to $\langle u_{x,f}\rangle_{\mathcal{W}}$ for `rejected' or `accepted' windows for $\overline{u}_{x,real}$~=~10~ms$^{-1}$ and $\mathrm{T}_{u,x,real}$ of (c) 5\% and (e) 35\%, respectively. Error bars represent 95\% confidence intervals.}
\label{fig:mean_velocities}
\end{figure}

Figure~\ref{fig:mean_velocities}a shows that mean velocities are consistently overestimated with increasing streamwise turbulence ($\mathrm{T}_{u,x}$). This overestimation remains below 10\% for $\mathrm{T_{u,x}}$ values up to 25\%, yet it increases to approximately 16\% when the streamwise turbulence intensity reaches 35\%. The observed bias is consistent across the whole range of mean velocities. \\

We found that error mechanism \textit{E3} was the key factor contributing to the observed bias in the mean velocity. In Figures~\ref{fig:mean_velocities}b,d, the frequency distributions of the window-averaged fluid velocity $\langle u_{x,f}\rangle_{\mathcal{W}}$, i.e., the window-averaged velocity of the true velocities of the bubbles impacting the probe tips (which are later analysed with the AWCC over the same time window) and the estimated pseudo-instantaneous velocity $u_{x,awcc}$ are compared for the cases of $\overline{u}_{x,real}$~=~10~ms$^{-1}$ and $\mathrm{T_{u,x,real}}$ of 5\% and 35\%, respectively. For $\langle u_{x,f}\rangle_{\mathcal{W}}$, we distinguish between `all' windows, and windows that were either `rejected' or `accepted' during the correlation-based filtering (Eq.~\ref{eq:correlation_filter} and ROC).\\

For small streamwise turbulence (Fig.~\ref{fig:mean_velocities}b), both accepted and rejected velocity estimations are distributed symmetrically around the imposed mean velocity $\overline{u}_{x,real}$~=~10~ms$^{-1}$. However, for $\mathrm{T_{u,x,real}}$~=~35\%, AWCC velocity estimations for windows with smaller flow velocities are more frequently rejected. This ultimately results in an over-representation of windows with higher flow velocities that explains the observed bias.
Despite causing a bias towards overestimation of mean velocities in case of large streamwise turbulence, we find that the correlation-based filter is very effective in removing numerous erroneous pseudo-instantaneous velocity estimations (see Fig.~\ref{fig:mean_velocities}e). The error of the accepted windows are largely within $\pm$~10\%.\\

With increasing transverse turbulence intensity $\mathrm{T_{u,y}}$ and $\mathrm{T_{u,z}}$, we identify a trend to slightly overestimate mean velocities in the order of up to $\sim 2$ \%, but only when turbulence intensities $\mathrm{T_{u,yz}}$ exceed 10\%. The observed overestimation is mainly attributed to error mechanisms \textit{E1} and \textit{E2}. At larger transverse turbulence intensities, bubbles may travel at larger or smaller angles $\theta$ with respect to the probe orientation. While bubbles travelling at angles $\alpha~<~\theta~<~0^\circ$ lie within the underestimation cone, bubbles travelling at angles $\theta~>0^\circ~$ or $\theta~<~\alpha$ are in the overestimation cone (Fig. \ref{fig:estimation_cones}a,b). Considering the specific geometry of the probe ($\alpha \approx 11^\circ$), the tendency to overestimate is more pronounced than the tendency to underestimate. This leads to the observed bias for large transverse turbulence intensity $\mathrm{T_{u,y}}$ and $\mathrm{T_{u,z}}$. This outcome aligns with expectations, given that the underestimation occurs within an internal cone, resulting in a more rapidly expanding probability space for the overestimation.\\

Additional tests suggest that other factors such as the void fraction and bubble size do not pose any significant impact on the mean velocity estimation for the range of flow properties studied (Table~\ref{tab:scenarios}). Factors such as the integral time scale or bubble size also resulted in negligible effects.

\subsection{Flow properties affecting the turbulence intensity estimation}

Turbulence intensity is one of the most complicated flow properties to be sampled in air-water flows. To estimate turbulence intensity, first a pseudo-instantaneous time series of velocities needs to be estimated from the raw signals. When doing so using AWCC, this is limited to an estimation of a ``pseudo"-instantaneous velocity because it represents the velocity of a group of bubbles. Here we systematically study how the estimation of turbulence is affected by the air-water flow properties.\\

\begin{figure}[!ht]
\makebox[\textwidth][c]{\includegraphics{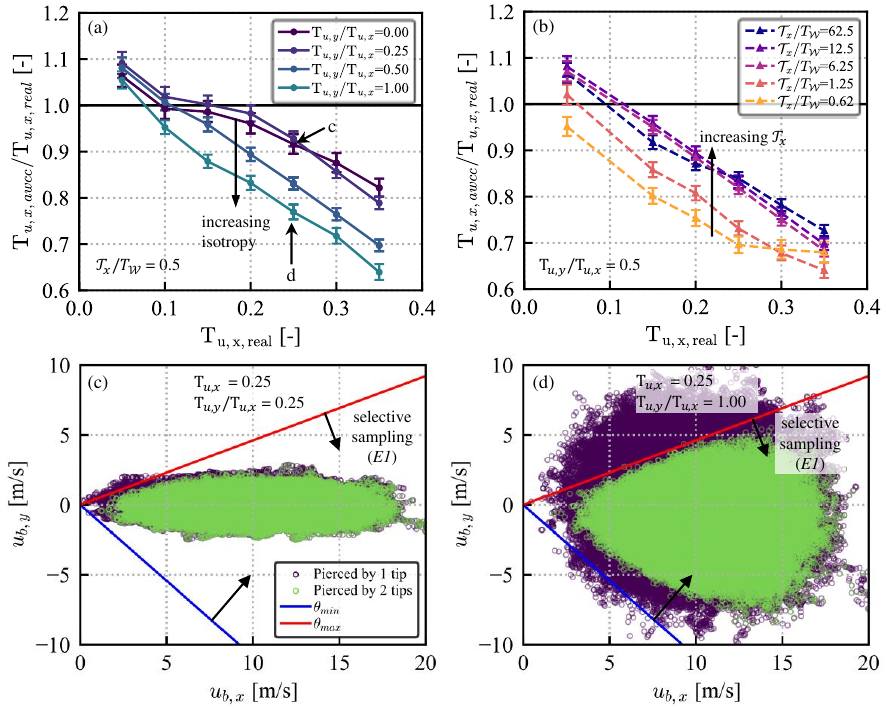}}
\caption{Effect of turbulent intensity on turbulence estimations, including effects of (a) anisotropy and (b) integral time scale of the flow. Direct illustration of \textit{E1} key error mechanism at low isotropic ($T_{u,y}/T_{u,x}$~=~0.25) (c) and fully isotropic ($T_{u,y}/T_{u,x}$~=~1) turbulence (d) with $T_{u,x}$~=~25\%. Error bars represent 95\% confidence intervals.}
\label{fig:TI_vsTI}
\end{figure}

Our analysis suggests that the prime factor affecting the estimation of streamwise turbulence intensity ($\mathrm{T}_{u,x,awcc}$) is the turbulence intensity of the flow itself. We consider streamwise turbulence intensities between 5\% and 35\%. Thereby we notice two effects: For small turbulence intensities $\mathrm{T}_{u,x}~\leq$~20$\%$ both under- and overestimations occur with errors mostly within $\pm$10\% of the actual $\mathrm{T}_{u,x,real}$, allowing a reasonably accurate turbulence estimation. However, as streamwise flow turbulence increases, AWCC estimations do not follow up linearly and the error in the estimated turbulence intensity becomes more significant. For larger turbulence intensities, $\mathrm{T}_{u,x,awcc}$ is always underestimated (Fig.~\ref{fig:TI_vsTI}\textit{a}). This underestimation can be as high as 30$\%$ of the actual $\mathrm{T}_{u,x,real}$, when the latter reaches values of 35$\%$. This means that in strong turbulence, direct estimation of turbulence intensity remains highly uncertain.\\

Additionally, we observe that the turbulence estimations are further affected by the anisotropy of the flow, i.e. by the ratio of transverse to streamwise turbulence intensity $T_{u,y}/T_{u,x}$ or $T_{u,z}/T_{u,x}$, hereinafter only referred to as  $T_{u,y}/T_{u,x}$. More isotropic turbulence exacerbates the underestimation at large $\mathrm{T}_{u,x,real}$.\\

Figures~\ref{fig:TI_vsTI}c-d presents the streamwise versus transverse velocities of bubbles pierced by either one or two tips for the case of $\mathrm{T}_{u,x,real}=0.25$ and anisotropy ratios $T_{u,y}/T_{u,x}$ of 0.25 and 1, respectively. Additionally, the minimum ($\theta_{min}$) and maximum ($\theta_{max}$) detection angles are indicated to illustrate the selective sampling bias due to the limit of detection ($\theta_{min}$ to $\theta_{max}$) formulated in Section \ref{sec:theory}, and highlight their importance. 
For $T_{u,y}/T_{u,x}$~=~0.25, most bubbles travel at angles that lie within the detection cone and the distribution of bubbles pierced by only one and two tips are comparable. However, with increasing $T_{u,y}/T_{u,x}$, bubbles with small $u_{b,x}$ and large $u_{b,y}$ may be detected by the leading tip while escaping from the trailing tip. As a consequence, bubbles with smaller streamwise velocities are less likely to be detected by both tips, resulting in the observed underestimation of turbulence intensities. This suggests that error mechanism \textit{E1}, i.e. the selective detection of bubbles, is the main driver for the larger underestimation with increasing turbulence isotropy.\\

We observe a second critical factor affecting the accurate estimation of turbulence in the flow, namely the integral time scale ($\mathcal{T}$). We test integral time scales between 0.005~s and 0.5~s. Considering the average window duration $T_{\mathcal{W}}$ of 0.008~s for the considered cases (Tab.~\ref{tab:scenarios}), this yields ratios $\mathcal{T}_x/T_{\mathcal{W}}$ between 0.625 and 62.5. Our results in Figure~\ref{fig:TI_vsTI}\textit{a} show that smaller integral time scales generally result in increasing underestimation of turbulence intensity. This is a direct effect of error mechanism \textit{E4}. As the ratio of integral time scale of the flow to average AWCC window duration $\mathcal{T}_x/T_{\mathcal{W}}$ becomes smaller, the most extreme turbulent fluctuations are more intensely smoothed out. For example, for $\mathrm{T_{u,y}}$~=~25\% and $\mathcal{T}_x/T_{\mathcal{W}}$~=~1.25 the underestimation amounts to roughly 25\%, which is comparable to the underestimation of 20\% found by \citet{Kramer2019b} for $\mathrm{T_{u,x}}$~=~20\% and $\mathcal{T}_x/T_{\mathcal{W}}\approx1$. However, with reducing $\mathcal{T}_x/T_{\mathcal{W}}$, the underestimation becomes more intense.\\

Another relevant factor identified in our analysis is related to air concentration. Our observations show that turbulence intensity is considerably underestimated for air concentrations $c < 5\%$ (for the base case). For air concentrations of 1\% the underestimation reaches up to 30\%. We identified \textit{E4} as main mechanism resulting in the observed trend. A smaller air concentration - for fixed remaining flow properties - results in a smaller bubble frequency. As a consequence, the average window duration $T_{\mathcal{W}}$ increases and the ratio of window duration to integral time scale decreases.\\

\begin{figure}[!ht]
\makebox[\textwidth][c]{\includegraphics{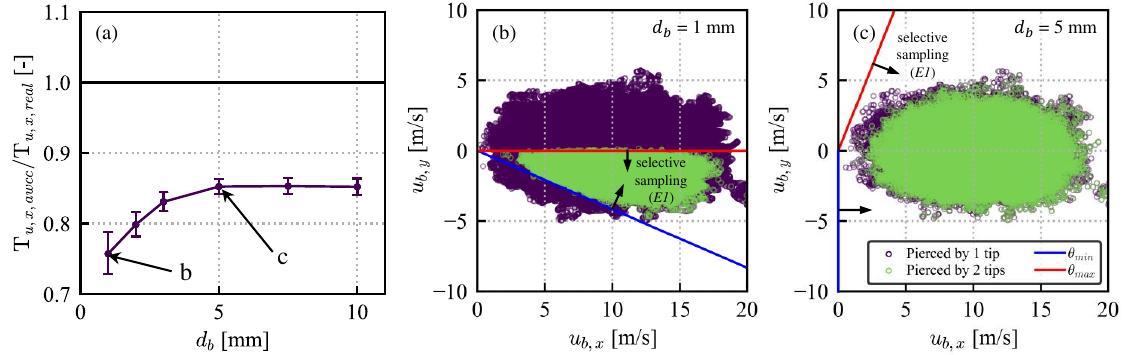}}
\caption{Effect of bubble size $d_b$ on turbulence intensity estimation for varying levels of turbulence anisotropy $\mathrm{T}_{u,y}/\mathrm{T}_{u,x}$ (a). Direct illustration of the effect of relative bubble size through mechanism \textit{E1} for a (b) small $d_b$~=~1~mm and (c) larger bubble size $d_b$~=~5~mm. Error bars represent 95\% confidence intervals.}
\label{fig:db_turbulent_intensity}
\end{figure}

Finally, we also examine how the bubble size with $d_b$ between 1~mm and 10~mm affect the estimation of turbulence intensities (Fig.~\ref{fig:db_turbulent_intensity}a). Decreasing the bubble size relative to the probe size increases the probability that bubbles are not pierced by both tips as direct consequence of mechanism \textit{E1}. This is consistent with the observation from Figure~\ref{fig:db_turbulent_intensity}a, that a bubble size smaller than the base case streamwise tip separation  $\Delta x$~=~5~mm results in increasing underestimation of turbulence intensity. As evident from Figure~\ref{fig:db_turbulent_intensity}b, a bubble size of $d_b$~=~1~mm results in a maximum bubble trajectory angle $\theta_{max}$ of 0$^\circ$, effectively preventing bubbles with a transverse velocity $u_{b,y}$ greater than 0 from interacting with both probe tips (\textit{E1}). This results in a reduced range of detected streamwise velocities and an underestimation of streamwise turbulence intensity by approximately 25\%. In contrast, for a larger $d_b$~=~5 the minimum and maximum angles of detection are not limiting for the considered flow conditions as evident in Figure~\ref{fig:db_turbulent_intensity}c.\\

\subsection{Effect of probe geometry and post-processing on mean velocity}
\label{sec:instrumentation_and_analysis}

Phase detection probes are often developed in research laboratories, which leads to differences in their design (Table \ref{tab:parameter_literature}). We investigate the effect of streamwise tip separation for a range of \mbox{$0.5~\mathrm{mm} \leq \Delta x \leq 10~\mathrm{mm}$}.
We discover a systematic overestimation of the mean velocity with increasing distance between leading and trailing tips (Fig~\ref{fig:delta_x_np}a). For a given bubble size $d_b$, streamwise tip separation $\Delta x$ and turbulence anisotropy ratio, bubbles travelling at smaller instantaneous streamwise velocities are more likely to miss the second tip due to transverse movements, than bubbles travelling at larger instantaneous streamwise velocity. Therefore, we have a detection bias towards larger velocities, based on mechanism \textit{E1}. Larger transverse velocity fluctuations due to a larger ratio of transverse to streamwise turbulence intensity $\mathrm{T}_{u,y}/\mathrm{T}_{u,x}$ exacerbate this trend, which is evident in Figure~\ref{fig:delta_x_np}a. The overestimation of the mean flow velocity may reach up to 20\% for streamwise tip separation of 10~mm and fully isotropic turbulence with $\mathrm{T}_{u,x,real}$~=~0.25. Nevertheless, for $\Delta x < 5$~mm, the error remains within $\pm$~10\%.\\

\subsection{Effect of probe geometry and post-processing on turbulence estimation}

\begin{figure}[ht]
\makebox[\textwidth][c]{\includegraphics{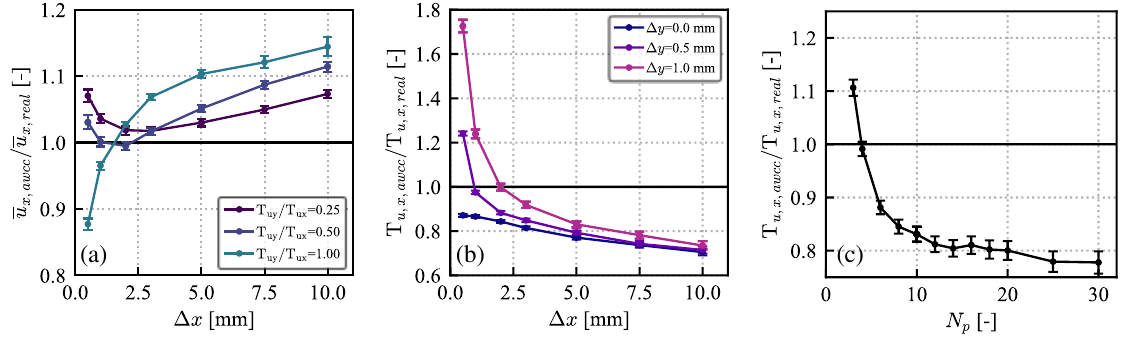}} \caption{Effect of streamwise tip separation $\Delta x$ on (a) mean velocity estimations for different levels of turbulence anisotropy, (b) turbulence intensity estimations for different lateral tip separations $\Delta y$. Smoothing effect of the number of particles per AWCC window ($N_p$) on the turbulence intensity estimations with AWCC and of the window averaged fluid velocity (c). Error bars represent 95\% confidence intervals.}
\label{fig:delta_x_np}
\end{figure}

Our results show that the streamwise tip separation $\Delta x$, but also the transverse separation $\Delta y$, affect turbulence intensity estimations (see Fig.~\ref{fig:delta_x_np}b). For probes with a streamwise alignment of the tips ($\Delta y$~=~0~mm), increasing $\Delta x$ results in increasing underestimation of turbulence intensity, reaching approx. 25\% for $\Delta x$~=~10~mm. For probes with a side-by-side design ($\Delta y > 0$), turbulence intensity may be overestimated when $\Delta x$ becomes smaller than $\approx$2~mm. Overestimations reach up to 170\% of the actual turbulence intensity for combinations of $\Delta x$~=~0.5~mm and $\Delta y$~=~1~mm, in combination with other base case parameters. This is consistent with larger overestimations expected from mechanism \textit{E2}. On the contrary, $\Delta x$ larger than 2~mm resulted in underestimations that tend to the case of zero transverse separation. The underestimations may be explained by two factors: (i) the smoothing effect of the window-averaging (\textit{E4}), and (ii) with increasing $\Delta x$, bubbles with large velocities are more likely to be detected, while slower bubbles are more likely to escape (due to traverse velocity fluctuations, i.e., mechanism \textit{E1}). This creates a bias towards larger mean velocities and underestimation of turbulence intensity as part of the histogram (slower velocities) is cut off by the selective sampling bias (\textit{E1}). We observed that increasingly isotropic turbulence enhances the observed trends. \\

Besides the streamwise and transverse tip spacing, we investigate the effect of the number of particles per averaging window $N_p$ on estimations of mean velocities and turbulence intensity. For the tested range of \mbox{$3\leq N_p \leq 30$,} the number of particles per window had no significant effect on mean velocity estimations. However, we observe that increasing the number of particles per averaging windows results in increasing underestimation (Fig.~\ref{fig:delta_x_np}c). This is in direct agreement with the findings of \citet{Kramer2020a}.

Finally, we also investigate the effect of sampling rates $f_s$ on the ability to estimate mean velocities and turbulence intensities.
Sampling rates applied in previous studies cover a wide range between 10~kHz and 1000~kHz. Our tests indicate that increasing the sampling rate above 20~kHz did not result in more accurate estimations of mean velocities and turbulent intensities. This is in agreement with previous studies \citep[e.g.][]{Felder2015}.

\section{Discussion}

\subsection{A regression model for the correction of baseline errors and how to use it (application)}
\label{sec:discussion:baseline_interpolation}

The previous analysis, for selected flow conditions around a base case (Table \ref{tab:scenarios}), showcases the order of magnitude of different errors and identifies key mechanisms (\textit{E1}-\textit{E4}, Section~\ref{sec:theory}) driving those inaccuracies. A direct quantification of errors (Section~\ref{sec:results}) is challenging, owing to the multi-dimensional, partially unknown parameter space, and due to the superposition of different error mechanisms. To enable a convenient estimation of the baseline error, we ran additional simulations comprising more than 19,000 parameters combinations using Monte Carlo sampling from the ranges identified in Table~\ref{tab:scenarios}. Based on this dataset of simulations \citep{Buergler2024a}, we trained a quantile random forest regression model \citep{Meinshausen2006}. The trained model allows to correct mean flow velocities and turbulence intensities obtained through AWCC. Corrected values are not only provided in terms of mean values, but through multiple quantiles, thereby, proving guidance on the uncertainty associated with the measurement and correction. \\

The required input parameters include the mean velocity, turbulence intensity, void fraction, and the Sauter mean diameter as a proxy for the bubble diameter, which are directly available from post-processing of phase-detection probe signals using AWCC. Further inputs include the streamwise and lateral tip separations $\Delta x$ and $\Delta y$, as well as the number of particles per window $N_p$. Details on the model and instructions for its application are available in \ref{app:correction_model}.\\

\begin{figure}[ht]
\makebox[\textwidth][c]{\includegraphics{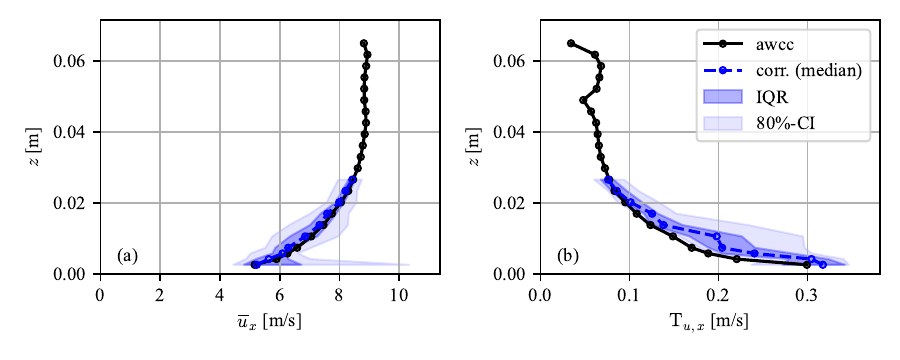}} \caption{Application of the regression model to quantify the corrected values (median), the interquartile range (IQR) and the 80\% confidence interval (CI) for a (a) mean velocity and (b) turbulence intensity profile measured by \citet{Buergler2023a} in gradually varied spillway flow characterized by an inclination angle of 50$^\circ$, hydraulic roughness of $k_s$ = 0.73~mm, and specific discharge of $q$~=~0.2~m$^2$s$^{-1}$. The application of the regression model is limited to lower half of the profile, where $c \leq 0.4$.}
\label{fig:rf_application}
\end{figure}

We demonstrate the application of the regression model for velocity and turbulence intensity profiles measured by \citet{Buergler2023a} in Figure \ref{fig:rf_application}. The trends for flow factors affecting the turbulence estimation have serious implications when measuring turbulence intensities in self-aerated boundary layer flows. Namely, the most adverse conditions -- a combination of large turbulence intensities with a typical anisotropy ratio of $\approx$~0.5 with small integral time scales, small air concentrations, and small bubble size -- occur close to the wall. Hence mean velocity and turbulence intensity measurements are least reliable where they would be most interesting to derive quantities like the shear velocity.\\

\subsubsection{Optimizing probe geometry}
\label{sec:discussion:physical}
Our results suggest that the accuracy of mean velocity and turbulence intensity estimations in bubbly flows strongly depends on the flow conditions, but also on the design of the phase-detection probes. In the presence of 3-D turbulence, a ratio of bubble size to tip separation distance smaller than 1 increases the risk that the measured velocity statistics are affected by selective sampling (\textit{E1}). This suggests that the tip-to-tip separation should be in the order of the prevailing bubble size. As the range of detectable bubble trajectory angles $\theta$ is symmetric around $\alpha$ -- the angle between the leading and trailing axis tip and the main flow direction -- an increasing ratio $\Delta y/\Delta x$ further increases the selective sampling bias. \\

Therefore, we recommend streamwise tip separations $\Delta x$ between 2~mm and 5~mm for typical transverse tip separations $\Delta y$ between 0.5~mm and 1~mm. With this, errors of mean velocity and turbulence intensity estimations are expected to be within $\pm$~10\% and $\pm$~20\%, respectively. \\

When the relative bubble size $d_b/\Delta y$ is smaller than 2, additional uncertainties are expected. While our results suggest that this may be compensated by reducing $\Delta y$, other issues -- such as as flow separation and a wake downstream of the leading tip \citep{Sene1984, Chanson1988} -- may become relevant as the streamwise separation approaches zero.\\

\subsection{Limitations}
\label{sec:discussion:limitations}
Our simulations of phase-detection probe measurements in bubbly flows are based on several simplifications. The probe tips are simplified as ideally small points, intrusive effects, such as slow down or deformation of bubbles, are not taken into account and spherical bubble shapes were assumed. In less ideal conditions, for example with varying bubble size and shape, a stronger decorrelation of the leading and trailing tip signals and therewith additional uncertainties are expected.

Other effects such as flow separation and a wake downstream of the leading tip \citep{Chanson1988}, as well as bubble-probe interactions \citep{Hohermuth2021b, Pagliara2023} are neglected. For regions close to the wall of turbulent boundary layers (small velocities and bubble size), \citet{Hohermuth2021b} observed a mean velocity bias of approximately -10\% in a comparison to LDA measurements due to intrusive effects. However, based on our findings an overestimation of mean velocities is expected for this region with large turbulence intensity. This potentially indicates that intrusive effects may actually be larger than reported by \citet{Hohermuth2021b}, but are to some degree compensated by baseline errors with opposing effects.\\
One question that arises is if the intrusive velocity bias correction scheme proposed by \citet{Hohermuth2021b} can be applied in addition to the correction model from this study. The fact that the intrusive velocity bias and the biases in estimated mean velocities presented in this study act in opposing directions, suggests that a correction of intrusive effects is still necessary after the correction of intrinsic errors. Since the correction scheme of \citet{Hohermuth2021b} was calibrated against LDA data without consideration of the biases presented herein, the correction scheme for intrusive effects may require a re-evaluation. \\

Finally, the correction model proposed herein should not be applied for phase-detection probe measurements if the flow conditions or probe characteristics lies outside of the ranges stated in Table~\ref{tab:scenarios}.   \\

\section{Conclusion}
Air-water flows have been sampled in civil engineering structures for nearly seven decades. However, studies on the performance of the specific instrumentation used in laboratory and field scale studies are scarce.
This study provides, for the first time, a complete and comprehensive description of a myriad of mechanisms that induce errors in the sampling of multiphase flows in spillways and tunnel flows. This is achieved by developing a tool for simulating phase-detection probe measurements, based on the Langevin equation and informed by real-world flow properties (Table \ref{tab:parameter_literature}).\\

This investigation represents the first uncertainty study taking into account 3-D turbulence for a large and complete range of flow properties. It is therefore able to reproduce realistic boundary layer characteristics inspired by real-word spillway data and can be directly used to estimate future experimental uncertainty.\\

Our investigation sheds light on the following points:
\begin{enumerate}
    \item We identify four key mechanisms that, combined, explain intrinsic baseline errors in the sampling of bubbly flows with phase-detection probes.
    More specifically, we delineate and quantify the selective sampling bias (\textit{E1}) and the decorrelation bias (\textit{E3}), both primarily attributed to 3-D turbulence, for the first time. Additionally, we quantify the previously described velocity bias due to transverse impacts of bubbles (\textit{E2}) and the averaging effect inherent to the AWCC processing algorithm (\textit{E4}).
    \item We quantify the magnitude of these errors and their interplay for the mean velocity and turbulence intensity estimations. Of paramount importance is the finding, that the least favorable conditions typically occur in the near wall region. The identified error mechanisms result in mean velocity error magnitudes of up to 15\%, which is comparable to the recently discussed intrusive velocity bias. Also turbulence intensities are substantially underestimated with increasing turbulence intensity of the flow itself and with increasing turbulence isotropy. In near-wall regions, the ascertained errors reach up to 40\% for typical phase-detection probe designs. This has serious implications when quantifying derived flow properties, such as the shear velocity, based on near-wall phase-detection probe measurements.
    \item While the errors induced by properties of the flow itself -- for example by high turbulence intensity -- can hardly be minimized, we show that the probe design is a key element in error minimization and that the optimal design depends on the bubble size. The presented, data-driven tool for bias correction of mean velocity and turbulence intensity estimations, but also the software for simulating phase-detection probe measurements, can be applied in future studies to improve the design of phase-detection probes for various flow conditions.
    \item Finally, we provide a novel, data-driven tool to correct for intrinsic errors occurring during phase-detection probe measurements and to quantify the associated uncertainties. This tool can be directly applied to past or future phase-detection probe measurements to correct mean velocities and turbulence intensities estimates. Thereby, this study contributes considerably to improved accuracy of dual-tip phase-detection probe measurements and the characterization of air-water flows in hydraulic structures.
\end{enumerate}

\section*{Appendix}

\appendix

\setcounter{figure}{0} 
\setcounter{subsection}{0} 
\renewcommand{\thefigure}{A\arabic{figure}}
\renewcommand{\thesubsection}{\Alph{section}}
\section{A regression model for the correction and uncertainty quantification of phase-detection probe measurements}\label{app:correction_model} 
The regression model for the bias correction and uncertainty quantification of phase-detection probe measurements is based on a large dataset of measurement errors containing more than 19,000 simulations \citep{Buergler2024a}. The simulation dataset represents a wide range of flow conditions, probe characteristics, and AWCC algorithm parameters as presented in Table~\ref{tab:scenarios}. All parameters were sampled uniformly over the stated range of values, except the integral time scale was sampled from a log-normal distribution. For each simulation, the measured mean velocity and turbulence intensity as well as the true underlying values are known. This allows to train a regression model also using the void fraction, the bubble diameter, the streamwise and lateral phase-detection probe tip separation, as well as the number of particles per window used for the AWCC algorithm as further inputs. The trained model is capable of correcting real-world phase-detection probe data for the intrinsic errors captured by the simulations.\\

\begin{figure}[H]
\makebox[\textwidth][c]{\includegraphics{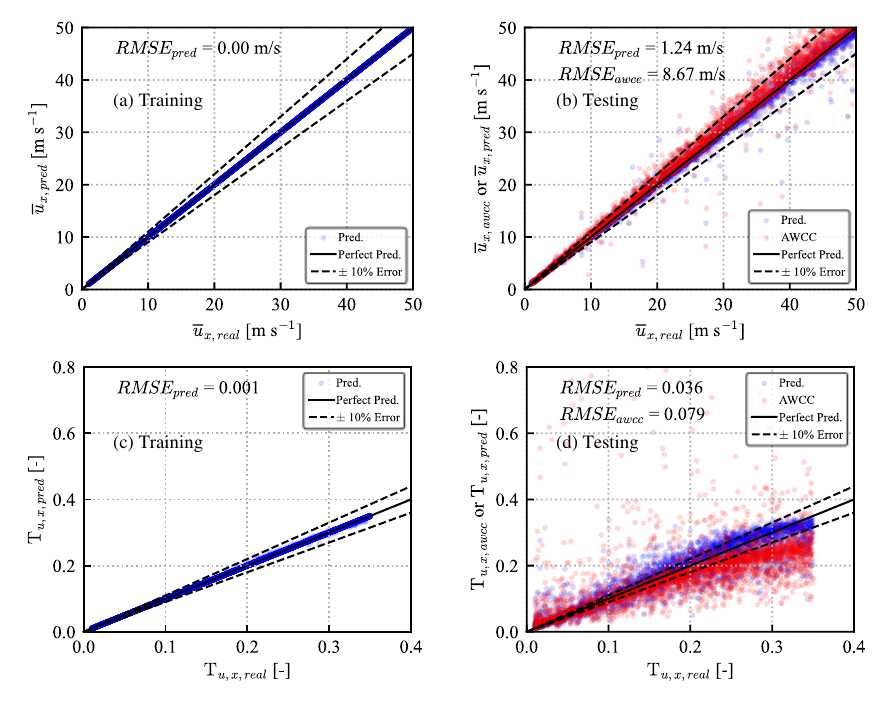}} \caption{Model predictions of mean velocities (a-b) and turbulence intensities (c-d) for the training and testing subsets. Mean velocities (b) and turbulence intensities (b) estimations resulting from the AWCC algorithm for the testing subset are presented for comparison.}
\label{fig:rf_validation}
\end{figure}

The regression model is based on a quantile regression forest model. A quantile regression forests is a non-parametric tool to estimate the conditional distribution of a target variable, suitable for high-dimensional input parameter space \citep{Meinshausen2006}. The model is implemented in Python, leveraging the existing `quantile-forest' Python package developed by \citet{Johnson2024}. Two separate quantile forest regression models were trained, one to correct the mean velocity and one to correct turbulence intensity estimations.
The data was split into training and testing sets containing 80\% and 20\% of the simulations, respectively. The training data was further split into 5 folds, which we used for hyperparameter tuning applying 300 iterations of a randomized grid search cross-validation. The optimized hyperparameters comprise number of trees, maximum tree depth and number of features considered for the best split.
The performance of the model prediction of mean velocity and turbulence intensity is illustrated in Figure~\ref{fig:rf_validation} for the training and testing set. Predicted values correspond to the median. For the testing dataset, the performance of the AWCC algorithm is presented for comparison. The prediction model is able to reduce the root mean square error for the mean velocity by a factor of 7 and for the turbulence intensity by a factor of 2.2, indicating a considerable improvement.
Besides median corrected values, the regression model also provides the 10- to 90-percentile and interquartile range.\\

Access to the code and detailed instructions on model usage, including input data format and parameter settings to facilitate a straightforward application of the regression model to dual-tip phase-detection probe measurements is provided at \url{https://gitlab.ethz.ch/vaw/public/pdp-ibc-uq} \citep{Buergler2024b}. 

\newpage

\section*{CRediT author statement}
\textbf{Matthias Bürgler}: Conceptualization, Methodology, Software, Validation, Formal analysis, Investigation, Writing - Original Draft, Visualization. \textbf{Daniel Valero}: Conceptualization, Formal analysis, Writing- Original draft preparation, Writing- Reviewing and Editing, Supervision, Funding acquisition. \textbf{Benjamin Hohermuth}: Conceptualization, Software, Writing- Reviewing and Editing, Supervision, Funding acquisition. \textbf{Robert M. Boes}: Writing- Reviewing and Editing, Supervision, Funding acquisition. \textbf{David F. Vetsch}: Writing- Reviewing and Editing, Supervision, Funding acquisition.

\section*{Funding}
The first and third author were supported by the Swiss National Science Foundation (SNSF) [grant number 197208].

\section*{Competing interests}
The authors declare that they have no conflict of interest.

\section*{Data availability}
The data analysed in this paper were produced with the software for simulating phase-detection probe measurements within bubbly flows \citep{Buergler2024c}, described in the code availability statement. The dataset of measurement errors containing more than 19,000 simulations used for training the regression model is available in \citet{Buergler2024a}.

\section*{Code availability}
The code of the software for simulating phase-detection probe measurements within bubbly flows is openly accessible at  \url{https://gitlab.ethz.ch/vaw/public/pdp-sim-tbf} \citep{Buergler2024c}. The code implementing the data-driven regression model is available at  \url{https://gitlab.ethz.ch/vaw/public/pdp-ibc-uq} \citep{Buergler2024b}. The regression model makes use of the existing `quantile-forest' Python package developed by \citet{Johnson2024}.


\begin{thebibliography}{00}

\bibitem[Beggs \& Brill(1973)]{Beggs1973} Beggs, D. H., \& Brill, J. P. 1973. A study of two-phase flow in inclined pipes.\textit{ Journal of Petroleum Technology}, 25(05), 607-617.

\bibitem[Benedict \& Gould(1973)]{Benedict1996} Benedict, L. H., \& Gould, R. D. 1996. Towards better uncertainty estimates for turbulence statistics.\textit{ Experiments in Fluids}, 22(2), 129-136.

\bibitem[Boes \& Hager(2003)]{Boes2003} Boes, R. M., \& Hager, W. H. 2003. Two-phase flow characteristics of stepped spillways.{\it Journal of Hydraulic Engineering}, 129(9), 661-670.

\bibitem[Buchhave et al.(1979)]{Buchhave1979} Buchhave, P., George Jr, W. K., \& Lumley, J. L. (1979). The measurement of turbulence with the laser-Doppler anemometer. {\it Annual Review of Fluid Mechanics}, 11(1), 443-503.

\bibitem[Bung (2011)]{Bung2011} Bung, D. B. 2011. Developing flow in skimming flow regime on embankment stepped spillways. {\it Journal of Hydraulic Research}, 49(5), 639-648.

\bibitem[Bung \& Valero(2016)]{Bung2016} Bung, D. B., \& Valero, D. 2016. Optical flow estimation in aerated flows. {\it Journal of Hydraulic Research}, 54(5), 575-580.

\bibitem[Bung \& Valero(2017)]{Bung2017} Bung, D. B., \& Valero, D. 2017. FlowCV-An open-source toolbox for computer vision applications in turbulent flows. In {\it Proceedings 37th IAHR World Congress} (pp. 5356-5365), Kuala Lumpur, Malaysia.

\bibitem[B\"urgler et al.(2022)]{Buergler2022} B\"urgler, M., Hohermuth, B., Vetsch, D. F., \& Boes, R. M. 2022. Comparison of Signal Processing Algorithms for Multi-Sensor Intrusive Phase-Detection Probes. In {\it Proceedings 39th IAHR World Congress} (pp. 5094-5103), Granada, Spain, International Association for Hydro-Environment Engineering and Research.

[dataset]\bibitem[B\"urgler et al.(2024a)]{Buergler2024a} B\"urgler, M., Valero, D., Hohermuth, B., Boes, R. M., \&  Vetsch, D. F. 2024a. Dataset for ``Uncertainties in Measurements of Bubbly Flows Using Phase-Detection Probes". ETH Zurich Research Collection. \url{https://doi.org/10.3929/ethz-b-000664463}.

\bibitem[B\"urgler et al.(2024b)]{Buergler2024b} B\"urgler, M., Valero, D., Hohermuth, B., Boes, R. M., \&  Vetsch, D. F. 2024b. Intrinsic Bias Correction and Uncertainty Quantification of Phase-Detection Probe Measurements.  ETH Zurich. \url{https://gitlab.ethz.ch/vaw/public/pdp-ibc-uq}.

\bibitem[B\"urgler et al.(2024c)]{Buergler2024c} B\"urgler, M., Valero, D., Hohermuth, B., Boes, R. M., \&  Vetsch, D. F. 2024c. Phase-Detection Probe Simulator for Turbulent Bubbly Flows. ETH Zurich. \url{https://gitlab.ethz.ch/vaw/public/pdp-sim-tbf}.


\bibitem[B\"urgler et al.(2023a)]{Buergler2023a} B\"urgler, M., Valero, D., Vetsch, D. F., Boes, R. M., \&  Hohermuth, B. 2023a. Air-Water Flow Measurements on a Large-Scale Spillway Chute. In {\it Proceedings 40th IAHR World Congress}, Vienna, Austria, International Association for Hydro-Environment Engineering and Research. 2119-2127.

\bibitem[B\"urgler et al.(2023b)]{Buergler2023b} B\"urgler, M., Vetsch, D. F., Boes, R. M., Hohermuth, B., \& Valero, D. 2023b. Effect of invert roughness on smooth spillway chute flow. In {\it Role of Dams and Reservoirs in a Successful Energy Transition} (pp. 687-695). CRC Press.

\bibitem[Cain(1978)]{Cain1978} Cain, P. 1978. {\it Measurements within self-aerated flow on a large spillway.} PhD thesis, Univ. of Canterbury, Christchurch, New Zealand.

\bibitem[Chanson(1988)]{Chanson1988} Chanson, H. 1988. {\it Study of air entrainment and aeration devices on spillway model.}, PhD thesis, Univ. of Canterbury, Christchurch, New Zealand.

\bibitem[Chanson(2013)]{Chanson2013} Chanson, H. 2013. Hydraulics of aerated flows: Qui pro quo?. {\it Journal of Hydraulic Research}, 51(3), 223-243.

\bibitem[Chanson(2020)]{Chanson2020} Chanson, H. 2020. On velocity estimations in highly aerated flows with dual-tip phase-detection probes-A commentary. {\it International Journal of Multiphase Flow}, 132, 103330.

\bibitem[Chanson \& Toombes(2002)]{Chanson2002} Chanson, H., \& Toombes, L. 2002. Experimental study of gas-liquid interfacial properties in a stepped cascade flow. {\it Environmental Fluid Mechanics}, 2, 241-263.

\bibitem[Crowe(2005)]{Crowe2005} Crowe, C. T. (Ed.). 2005. Multiphase Flow Handbook (Vol. 59). CRC press.

\bibitem[Cui et al.(2022)]{Cui2022} Cui, H., Felder, S., \& Kramer, M. 2022. Multilayer velocity model predicting flow resistance of aerated flows down grass-lined spillway. {\it Journal of Hydraulic Engineering}, 148(10), 06022014.

\bibitem[Felder \& Chanson(2015)]{Felder2015} Felder, S., \& Chanson, H. 2015. Phase-detection probe measurements in high-velocity free-surface flows including a discussion of key sampling parameters. {\it Experimental Thermal and Fluid Science}, 61, 66-78.

\bibitem[Felder \& Chanson(2016)]{Felder2016} Felder, S., \& Chanson, H. 2016. Air–water flow characteristics in high-velocity free-surface flows with 50\% void fraction. {\it International Journal of Multiphase Flow}, 85, 186-195.

\bibitem[Felder \& Chanson(2017)]{Felder2017a} Felder, S., \& Chanson, H. 2017. Scale effects in microscopic air-water flow properties in high-velocity free-surface flows. {\it Experimental Thermal and Fluid Science}, 83, 19-36.

\bibitem[Felder et al.(2019)]{Felder2019} Felder, S., Hohermuth, B., \& Boes, R. M. 2019. High-velocity air-water flows downstream of sluice gates including selection of optimum phase-detection probe. {\it International Journal of Multiphase Flow}, 116, 203-220.

\bibitem[Felder \& Pfister(2017)]{Felder2017b} Felder, S., \& Pfister, M. 2017. Comparative analyses of phase-detective intrusive probes in high-velocity air–water flows. {\it International Journal of Multiphase Flow}, 90, 88-101.

\bibitem[Felder et al.(2023)]{Felder2023} Felder, S., Severi, A., \& Kramer, M. 2023. Self-Aeration and Flow Resistance in High-Velocity Flows Down Spillways with Microrough Inverts. {\it Journal of Hydraulic Engineering}, 149(6), 04023011.

\bibitem[Hohermuth et al.(2021a)]{Hohermuth2021a} Hohermuth, B., Boes, R. M., \& Felder, S. 2021. High-velocity air–water flow measurements in a prototype tunnel chute: Scaling of void fraction and interfacial velocity. {\it Journal of Hydraulic Engineering}, 147(11), 04021044.

\bibitem[Hohermuth et al.(2021b)]{Hohermuth2021b} Hohermuth, B., Kramer, M., Felder, S., \& Valero, D. 2021. Velocity bias in intrusive gas-liquid flow measurements. {\it Nature Communications}, 12(1), 4123.

\bibitem[Hornung et al.(1995)]{Hornung1995} Hornung, H. G., Willert, C., \& Turner, S. 1995. The flow field downstream of a hydraulic jump. {\it Journal of Fluid Mechanics}, 287, 299-316.

\bibitem[Johnson (2024)]{Johnson2024} Johnson, R. A. 2024. quantile-forest: A Python Package for Quantile Regression Forests. {\it Journal of Open Source Software}, 9(93), 5976.

\bibitem[Jones \& Delhaye(1976)]{Jones1976} Jones Jr, O. C., \& Delhaye, J. M. 1976. Transient and statistical measurement techniques for two-phase flows: a critical review. {\it International Journal of Multiphase Flow}, 3(2), 89-116.

\bibitem[Kramer(2019)]{Kramer2019a} Kramer, M. 2019. Particle size distributions in turbulent air-water flows. In {\it E-Proceedings of the 38th IAHR World Congress} (pp. 5722-5731).

\bibitem[Kramer et al.(2019)]{Kramer2019b} Kramer, M., Valero, D., Chanson, H., \& Bung, D. B. 2019. Towards reliable turbulence estimations with phase-detection probes: an adaptive window cross-correlation technique. {\it Experiments in Fluids}, 60(1), 2.

\bibitem[Kramer et al.(2020)]{Kramer2020a} Kramer, M., Hohermuth, B., Valero, D., \& Felder, S. 2020. Best practices for velocity estimations in highly aerated flows with dual-tip phase-detection probes. {\it International Journal of Multiphase Flow}, 126, 103228.

\bibitem[Kramer et al.(2021)]{Kramer2021} Kramer, M., Hohermuth, B., Valero, D., \& Felder, S. 2021. On velocity estimations in highly aerated flows with dual-tip phase-detection probes -- closure. {\it International Journal of Multiphase Flow}, 134, 103475.

\bibitem[Kramer \& Valero(2020)]{Kramer2020b} Kramer, M., \& Valero, D. 2020. Turbulence and self-similarity in highly aerated shear flows: The stable hydraulic jump. {\it International Journal of Multiphase Flow}, 129, 103316.

\bibitem[Matos et al.(2002)]{Matos2002} Matos, J., Frizell, K. H., André, S., \& Frizell, K. W. 2002. On the performance of velocity measurement techniques in air-water flows. In {\it Hydraulic Measurements and Experimental Methods 2002} (pp. 1-11).

\bibitem[Meinshausen \& Ridgeway(2006)]{Meinshausen2006} Meinshausen, N., \& Ridgeway, G. 2006. Quantile regression forests. {\it Journal of Machine Learning Research}, 7(6).

\bibitem[Meireles et al.(2012)]{Meireles2012} Meireles, I., Renna, F., Matos, J., \& Bombardelli, F. 2012. Skimming, nonaerated flow on stepped spillways over roller compacted concrete dams. {\it Journal of Hydraulic Engineering}, 138(10), 870-877.

\bibitem[Nikora \& Goring(1998)]{Nikora1998} Nikora, V. I., \& Goring, D. G. 1998. ADV measurements of turbulence: Can we improve their interpretation?. {\it Journal of Hydraulic Engineering}, 124(6), 630-634.

\bibitem[Nina et al.(2022)]{Nina2022} Nina, Y. A., Shi, R., Wüthrich, D.,\& Chanson, H. 2022. Intrusive and non-intrusive two-phase air-water measurements on stepped spillways: A physical study. {\it Experimental Thermal and Fluid Science}, 131, 110545.

\bibitem[Pagliara et al.(2023)]{Pagliara2023} Pagliara, S., Felder, S., Boes, R. M., \& Hohermuth, B. 2023. Intrusive effects of dual-tip conductivity probes on bubble measurements in a wide velocity range. {\it International Journal of Multiphase Flow}, 104660.

\bibitem[Pope(2000)]{Pope2000} Pope, S. B. 2000. {\it Turbulent flows}. Cambridge university press.

\bibitem[Sene(1984)]{Sene1984} Sene, K. J. 1984. Aspects of bubbly two-phase flow. {\it PhD thesis}, Trinity College, Cambridge, U.K.

\bibitem[Straub \& Anderson(1958)]{Straub1958} Straub, L. G., \& Anderson, A. G. 1958. Experiments on self-aerated flow in open channels. {\it Journal of the hydraulics division}, 84(7), 1-35.

\bibitem[Thorwarth(2008)]{Thorwarth2008} Thorwarth, J., 2008. Hydraulics on Pooled Stepped Chutes - Self Induced Unsteady Flow and Energy Dissipation (in German). RWTH Aachen, Germany. PhD thesis.

\bibitem[Vahedifard et al.(2017)]{Vahedifard2017} Vahedifard, F., AghaKouchak, A., Ragno, E., Shahrokhabadi, S., \& Mallakpour, I. 2017. Lessons from the Oroville dam. {\it Science}, 355(6330), 1139-1140.

\bibitem[Valero et al.(2022)]{Valero2022} Valero, D., Belay, B. S., Moreno-Rodenas, A., Kramer, M., \& Franca, M. J. 2022. The key role of surface tension in the transport and quantification of plastic pollution in rivers.{\it Water Research}, 226, 119078.

\bibitem[Valero et al.(2020)]{Valero2020} Valero, D., Chanson, H., \& Bung, D.B. 2020. Robust estimators for free surface turbulence characterization: A stepped spillway application. {\it Flow Measurement and Instrumentation}, 76, 101809.

\bibitem[Valero et al.(2019)]{Valero2019} Valero, D., Kramer, M., Bung, D.B., \& Chanson, H. 2019. A stochastic bubble generator for air water flow research. In {\it E-Proceedings of the 38th IAHR World Congress}. Panama City, Panama, pp. 5714–5721.

\bibitem[Velte et al.(2014)]{Velte2014} Velte, C. M., George, W. K., \& Buchhave, P. 2014. Estimation of burst-mode LDA power spectra. {\it Experiments in Fluids}, 55, 1-20.

\bibitem[Vernier \& Delhaye(1968)]{Vernier1968} Vernier, P., \& Delhaye, J. M. 1968. General two-phase flow equations applied to the thermohydrodynamics of boiling nuclear reactor.{\it Energie primaire}, 4(1), 3-43.

\bibitem[Wahl(2003)]{Wahl2003} Wahl, T. L. 2003. Discussion of “Despiking acoustic doppler velocimeter data” by Derek G. Goring and Vladimir I. Nikora. {\it Journal of Hydraulic Engineering}, 129(6), 484-487.

\bibitem[Wahl et al.(2019)]{Wahl2019} Wahl, T. L., Frizell, K. W., \& Falvey, H. T. 2019. Uplift pressures below spillway chute slabs at unvented open offset joints. {\it Journal of Hydraulic Engineering}, 145(11), 04019039.

\bibitem[Wang et al.(2018)]{Wang2018} Wang, D., Liu, Y., \& Talley, J. D. 2018. Numerical evaluation of the uncertainty of double-sensor conductivity probe for bubbly flow measurement. {\it International Journal of Multiphase Flow}, 107, 51-66.

\bibitem[Wang et al.(2021)]{Wang2021b} Wang, K., Tang, R., Bai, R., \& Wang, H. 2021. Evaluating phase-detection-based approaches for interfacial velocity and turbulence intensity estimation in a highly-aerated hydraulic jump. {\it Flow Measurement and Instrumentation}, 81, 102045.

\bibitem[Welch(1967)]{Welch1967} Welch, P. 1967. The use of fast Fourier transform for the estimation of power spectra: a method based on time averaging over short, modified periodograms. {\it IEEE Transactions on Audio and Electroacoustics, 15(2)}, 70-73.

\bibitem[Zhang \& Chanson(2016)]{Zhang2016} Zhang, G., \& Chanson, H. 2016. Hydraulics of the developing flow region of stepped spillways. I: Physical modeling and boundary layer development. {\it Journal of Hydraulic Engineering}, 142(7), 04016015.

\bibitem[Zhang \& Chanson(2018)]{Zhang2018} Zhang, G., \& Chanson, H. 2018. Application of local optical flow methods to high-velocity free-surface flows: Validation and application to stepped chutes. {\it Experimental Thermal and Fluid Science}, 90, 186-199.

\end{thebibliography}
\end{document}